\documentclass[10pt]{article}
\usepackage{OSAmeet}
\def\bfR{{\bf R}}
\def\bfZ{{\bf Z}}

\def\defi{\stackrel{\rm def}{=}}
\begin{document}
\title{Two Means of Compensating Fiber Nonlinearity\\
Using Optical Phase Conjugation}
\author{Haiqing Wei$^*$ and David V. Plant}
\address{Department of Electrical and Computer Engineering\\
McGill University, Montreal, Canada H3A-2A6}
\email{$^*$hwei1@po-box.mcgill.ca} \pagestyle{plain}
\begin{abstract}
Two fiber lines may compensate each other for nonlinearity with
the help of optical phase conjugation. The pair of fiber lines and
the optical signals in them may be either mirror-symmetric or
translationally symmetric about the conjugator.
\end{abstract}

\begin{figure}[h]
\begin{center}
\begin{minipage}{67mm}
\scalebox{.43}{\includegraphics{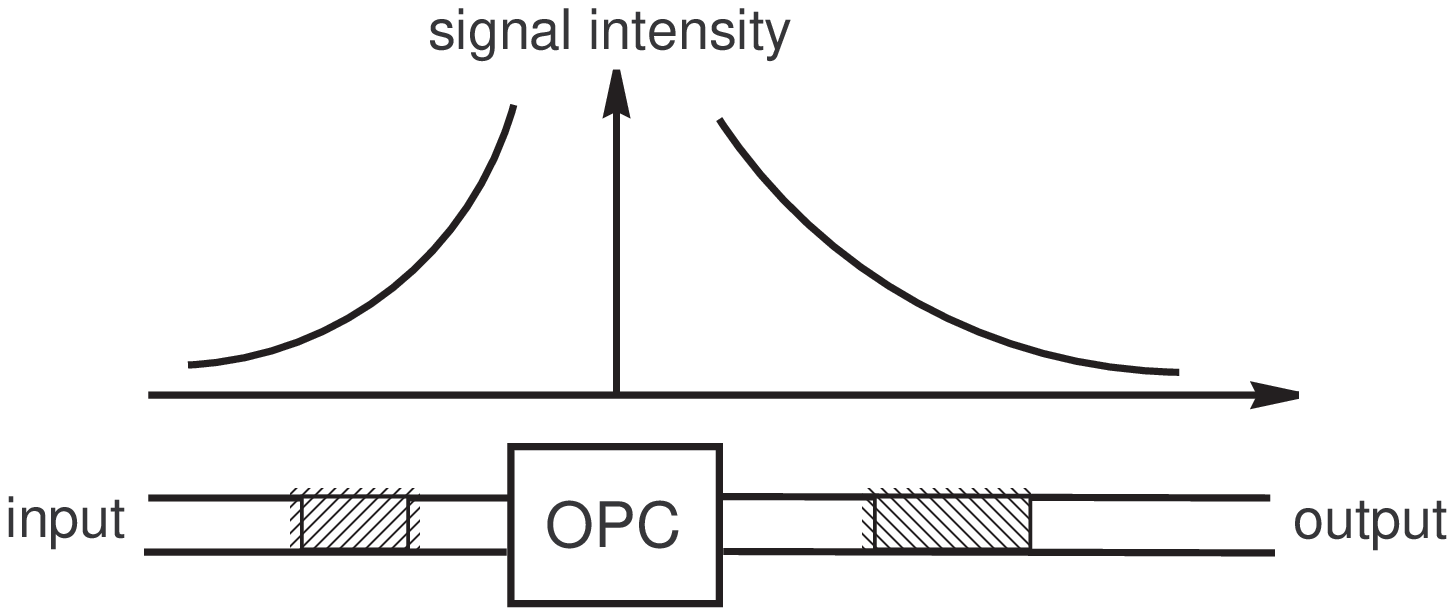}}
\end{minipage}
\hspace{1mm}
\begin{minipage}{67mm}
\scalebox{.43}{\includegraphics{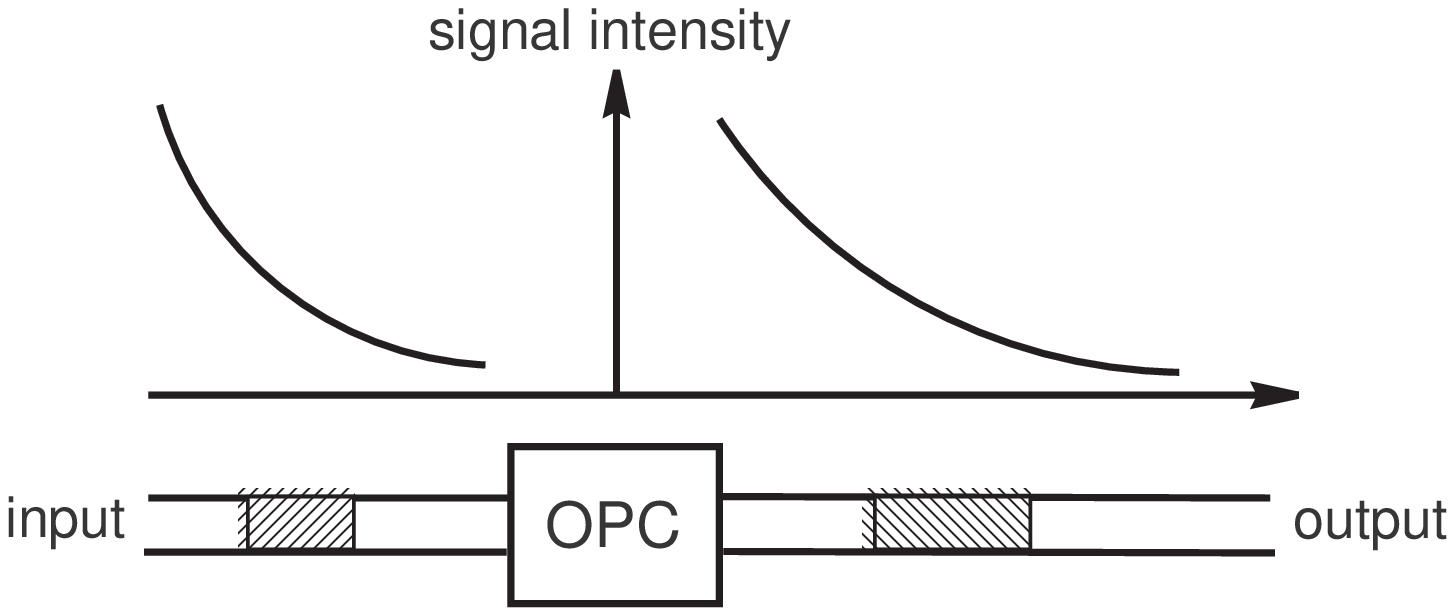}}
\end{minipage}
\end{center}\vspace{-6mm}
\caption{\label{2types}Two types of fiber arrangement for
nonlinearity compensation with OPC: mirror-symmetric (left) and
translationally symmetric (right).}
\end{figure}

The growing demand of higher capacity over longer transmission
distances has seen fiber nonlinearity as a major limiting factor
in modern optical transmission systems \cite{Forghieri97,Mitra01}.
Among the methods under investigation, nonlinearity compensation
using optical phase conjugation (OPC) has emerged as an effective
means of suppressing the nonlinear impairments
\cite{Watanabe96,Brener00,Wei03really}. This paper shall discuss
two types of fiber arrangement with respect to the OPC, as shown
in Fig.\ref{2types}. In one type of arrangement, the fiber
parameters and the signal intensity are in scaled mirror symmetry
about the OPC. While the other type is characterized by scaled
translational symmetry.

A mirror-symmetric link may consist of a fiber line on the left
stretching from $z=-L/R$ to $z=0$, $L>0$, $R>0$, followed by an
OPC, then a fiber line on the right stretching from $z=0$ to
$z=L$. The two fiber lines may carry wavelength-division
multiplexed (WDM) signals
$\sum_nA_n(z,t)\exp\left[i\int^z\beta(\zeta,\omega_n)d\zeta
-i\omega_nt\right]$ and
$\sum_nA'_n(z,t)\exp\left[i\int^zq(\zeta,\omega'_n)d\zeta
-i\omega'_nt\right]$ respectively, where $\forall n\in\bfZ$,
$\omega_n=\omega_0+n\Delta$ and $\omega'_n=\omega'_0+n\Delta$ are
the center frequencies of the WDM channels, $\Delta>0$, $\omega_0$
is not necessarily equal to $\omega'_0$, $A_n$ and $A'_n$ are the
slow-varying envelopes, while $\beta(z,\omega)$ and $q(z,\omega)$
are the $z$-dependent propagation constants. Being neglected is
the random polarization-mode dispersion (PMD) effect. And for
mathematical simplicity, all optical signals are assumed
co-linearly polarized in the fibers. Define short-hand notations
$\beta^{(k)}(z,\omega)\defi\frac{\partial^k} {\partial\omega^k}
\beta(z,\omega)$, and $q^{(k)}(z,\omega)\defi \frac{\partial^k}
{\partial\omega^k}q(z,\omega)$, $\forall k>0$. The dynamics of
signal propagation in the two fiber lines is governed by two
groups of coupled partial differential equations
\cite{Wei03really,Shen84,Agrawal95} respectively,
\begin{eqnarray}
\frac{\partial A_n}{\partial
z}-i\beta_{2n}\left(z,i\frac{\partial}{\partial
t}\right)A_n+\frac{\alpha_n(z)}{2}A_n=i\gamma(z)\sum_k
\sum_lA_kA_lA_{k+l-n}^*\exp[i\theta_{kln}(z)]+
\sum_mg_{mn}(z)|A_m|^2A_n,\label{dA}\\
\frac{\partial A'_n}{\partial
z}-iq_{2n}\left(z,i\frac{\partial}{\partial
t}\right)A'_n+\frac{\alpha'_n(z)}{2}A'_n=i\gamma'(z)\sum_k
\sum_lA'_kA'_lA'^*_{k+l-n}\exp[i\theta'_{kln}(z)]+
\sum_mg'_{mn}(z)|A'_m|^2A'_n,\label{dA'}
\end{eqnarray}
$\forall n\in\bfZ$, where for the first fiber line, $\alpha_n$ is
the attenuation coefficient around $\omega_n$, $\gamma$ is the
Kerr nonlinear coefficient, $g_{mn}$ is the Raman coupling
coefficient from the $m$th to the $n$th channels,
$\theta_{kln}(z)\defi\int^z[\beta(\zeta,\omega_k)+
\beta(\zeta,\omega_l)-\beta(\zeta,\omega_{k+l-n})-
\beta(\zeta,\omega_n)]d\zeta$ is the phase mismatch among the
mixing waves, and the functional operator $\beta_{2n}$ is defined
as $\beta_{2n}(z,i\frac{\partial}{\partial
t})\defi\sum_{k=1}^{+\infty}\frac{1}{k!}\beta^{(k)}(z,\omega_n)
\left(i\frac{\partial}{\partial t}\right)^k-\beta^{(1)}
(z,\omega_0) \left(i\frac{\partial}{\partial t}\right)$. The
parameters $\alpha'$, $\gamma'$, $g'_{mn}$, $\theta'_{kln}$ and
the operator $q_{2n}(z,i\frac{\partial}{\partial t})$ are
similarly defined for the second fiber line. It is an easy
exercise to show that equations (\ref{dA}) reduce to (\ref{dA'}),
when the parameters satisfy the following rules of correspondence,
\begin{eqnarray}
\beta^{(2)}(-z,\omega_0+\omega)&=&Rq^{(2)}(Rz,\omega'_0-\omega),
~~\forall\omega\in\bfR,\label{beta2q}\\
\alpha_n(-z)&=&-R\alpha'_{-n}(Rz),~~\forall
n\in\bfZ,\label{alpha'}\\
\gamma(-z)&=&R\gamma'(Rz)|C|^{-2},\label{gamma'}\\
g_{mn}(-z)&=&-Rg'_{-m,-n}(Rz)|C|^{-2},~~\forall m,n\in\bfZ,
\label{gmn'}
\end{eqnarray}
$\forall z\in[0,L/R]$, $C\neq 0$ being a constant, and the
envelope functions are related as $A_n(-z,t)=CA'^*_{-n}(Rz,t)$,
$\forall n\in\bfZ$. Physically, it says that the two fiber lines
compensate each other for dispersion and nonlinearity. Optical
signals $A_n(-L/R,t)$, $n\in\bfZ$, entering the first fiber line
may be dispersed and nonlinearly distorted to become $A_n(0,t)$,
$n\in\bfZ$, which are converted into
$A'_n(0,t)=A^*_{-n}(0,t)/C^*$, $n\in\bfZ$, by the OPC. The second
fiber line will then propagate the optical signals in a reversed
manner with respect to the first. The final outputs signals
$A'_n(L,t)=A^*_{-n}(-L/R,t)/C^*$, $n\in\bfZ$, are exact replicas
of the initial signals up to complex conjugation. It is noted that
parts of one fiber line would amplify light in correspondence to
the attenuation in parts of the other, and vice versa. A specialty
fiber may be chosen with parameters satisfying equations
(\ref{beta2q},\ref{gamma'},\ref{gmn'}) to be the scaled mirror
image of a transmission fiber which usually attenuates light. At
the same time, erbium doping or Raman pumping should be employed
to obtain the gain specified by (\ref{alpha'}).

A link with translational symmetry could be constructed to cancel
weak nonlinearities up to the first order perturbation. Consider
two fiber lines with opposite Kerr and Raman nonlinear
coefficients but identical linear parameters. If (\ref{dA}) with
$z\in[-L,0]$ describe the signal propagation in one fiber line,
then the signal dynamics in the other would be governed by similar
equations with negative $\gamma$ and $g$ coefficients,
\begin{eqnarray}
\frac{\partial B_n}{\partial
z}-i\beta_{2n}\left(z-L,i\frac{\partial}{\partial
t}\right)B_n+\frac{\alpha_n(z-L)}{2}B_n=-i\gamma(z-L)\sum_k
\sum_lB_kB_lB_{k+l-n}^*\exp[i\theta_{kln}(z-L)]\nonumber\\
-\sum_mg_{mn}(z-L)|B_m|^2B_n,~~0\le z\le L,\label{dB}
\end{eqnarray}
which take the input $B_n(0,t)$, $n\in\bfZ$ and give the output
$B_n(L,t)$, $n\in\bfZ$. When the signal intensity is not very
high, so that the nonlinearity is weak and treated with
perturbation theory, the output from each fiber line is a linearly
dispersed version of the input, plus nonlinear distortions
expanded in power series of the $\gamma$ and $g$ coefficients. By
neglecting the higher order powers and keeping only the terms
linear in $\gamma$ or $g$, it can be seen that the two fiber lines
induce opposite nonlinear distortions to otherwise the same,
linearly dispersed signals. If the overall dispersion of each line
is compensated to zero and the signal loss is made up by a linear
optical amplifier, then the two lines in cascade would comprise a
transmission line with fiber nonlinearity annihilated up to the
first order perturbation. The problem is that an optical fiber
with negative nonlinear coefficients does not exist naturally.
Fortunately, it can be simulated by a regular fiber with the help
of OPC. Take a regular fiber with parameters
$(q,\alpha',\gamma',g')$ that satisfy,
\begin{eqnarray}
q^{(2)}(z,\omega'_0+\omega)&=&-R\beta^{(2)}(Rz-L,\omega_0-
\omega),~~\forall\omega\in\bfR,\label{beta2q2}\\
\alpha'_n(z)&=&R\alpha_{-n}(Rz-L),~~\forall
n\in\bfZ,\label{alpha'2}\\
\gamma'(z)&=&R\gamma(Rz-L)|C|^{-2},\label{gamma'2}\\
g'_{mn}(z)&=&-Rg_{-m,-n}(Rz-L)|C|^{-2},~~\forall m,n\in\bfZ,
\label{gmn'2}
\end{eqnarray}
$\forall z\in[0,L/R]$. The signal propagation in the regular fiber
is then governed by,
\begin{equation}
\frac{\partial B'_n}{\partial
z}-iq_{2n}\left(z,i\frac{\partial}{\partial
t}\right)B'_n+\frac{\alpha'_n(z)}{2}B'_n=i\gamma'(z)\sum_k
\sum_lB'_kB'_lB'^*_{k+l-n}\exp[i\theta'_{kln}(z)]+
\sum_mg'_{mn}(z)|B'_m|^2B'_n,\label{dB'}
\end{equation}
$\forall z\in[0,L/R]$, which are solved by
$B'_n(z,t)=CB^*_{-n}(Rz,t)$, $n\in\bfZ$, and turn the input
$B'_n(0,t)=CB^*_{-n}(0,t)$, $n\in\bfZ$, into the output
$B'_n(L/R,t)=CB^*_{-n}(L,t)$, $n\in\bfZ$. The regular fiber
equipped with OPC at its two ends takes the input $B_n(0,t)$,
$n\in\bfZ$ and gives the output $B_n(0,t)$, $n\in\bfZ$. That
fulfils the function of the fictitious fiber with negative
nonlinearity. The OPC at the output end of the regular fiber may
be omitted in practice, as most applications would not
differentiate between a signal and its conjugate.

\begin{figure}[h]
\begin{center}
\begin{minipage}{78mm}
\scalebox{.37}{\includegraphics{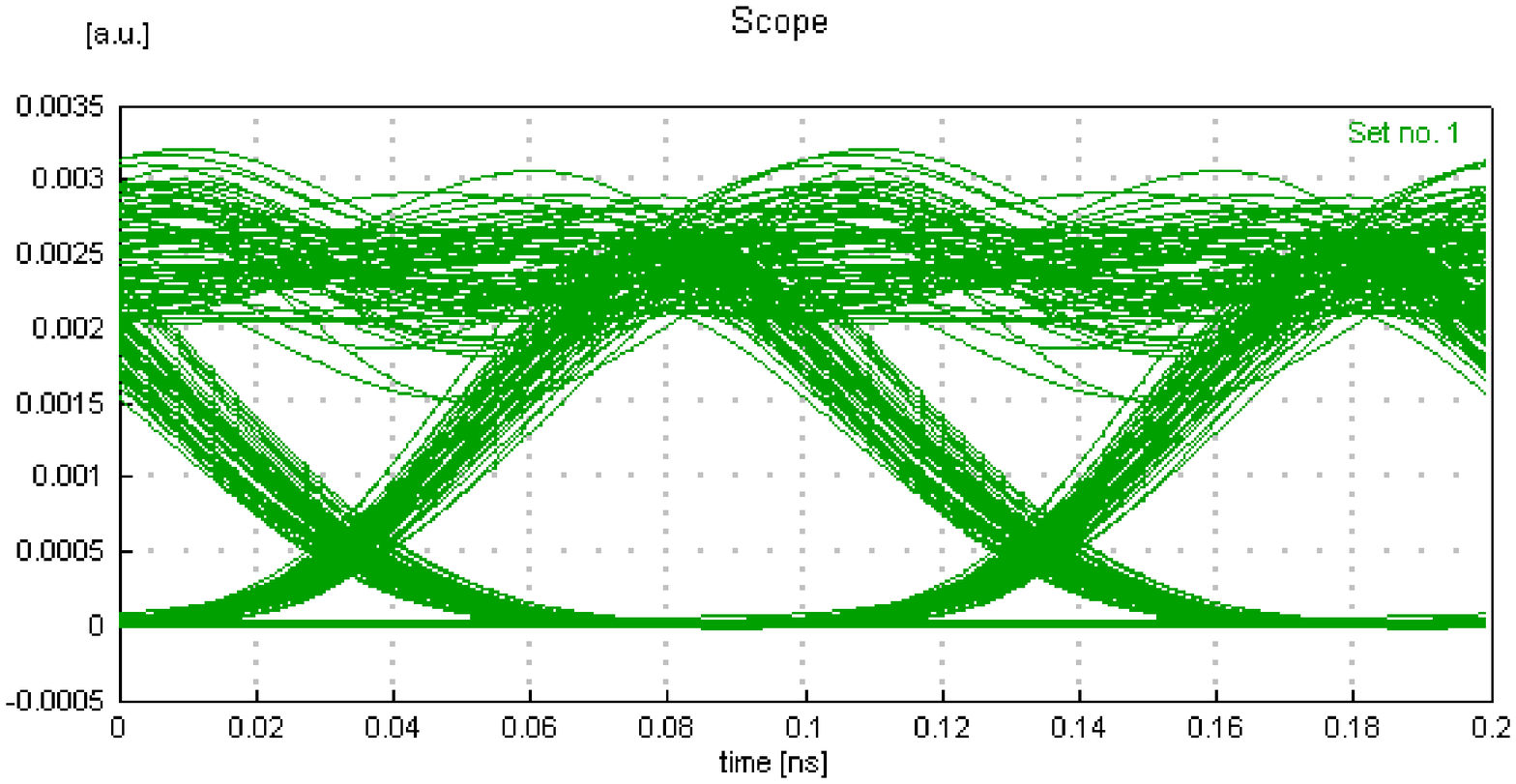}}
\end{minipage}
\begin{minipage}{78mm}
\scalebox{.37}{\includegraphics{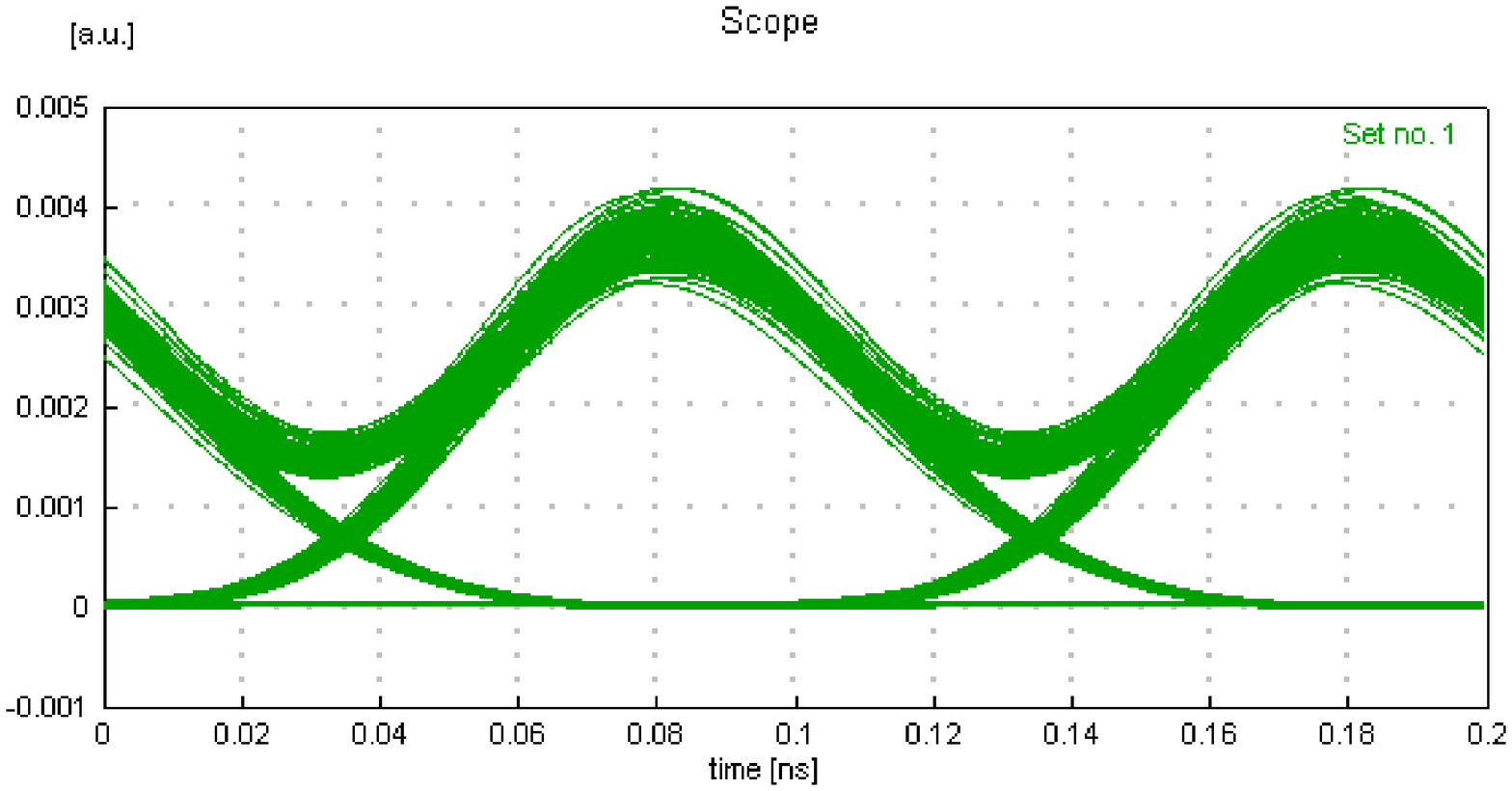}}
\end{minipage}
\end{center}\vspace{-6mm}
\caption{\label{simu4transl}Received signals at the end of the
translational link, without (left) and with (right) the OPC.}
\end{figure}

It is noted that each fiber line on one side of the OPC is not
necessarily one fiber span, and the signal intensity does not have
to evolve monotonically either. Both methods work fine when each
side of the OPC consists of multiple fiber spans with optical
amplifiers boosting the signal power, although the added noise
makes perfect nonlinearity compensation impossible. Using a
commercial software, computer simulations have been carried out to
test the proposed methods of nonlinearity compensation. For the
mirror setup, the test link consists of a specialty fiber, an OPC,
and a transmission fiber $200$ km long, with loss $\alpha=0.2$
dB/km, dispersion $D=-8$ ps/nm/km, dispersion slope $S=0.08$
ps/nm$^2$/km, effective mode area $A_{\rm eff}=50$ $\mu$m$^2$,
Kerr and Raman coefficients that are typical of silica glass. The
specialty fiber is made of the same material, but with parameters
$(\alpha',D',S')=20\times(-\alpha,D,-S)$ and $A'_{\rm eff}=12.5$
$\mu$m$^2$. The nonlinearity of the specialty fiber can be
switched on and off. Amplifier noise is added at the two ends of
the link. The input are four WDM channels spaced by $100$ GHz,
return-to-zero modulated at $10$ Gb/s with $33\%$ duty. The pulses
peak at $100$ mW when entering the transmission fiber. For a WDM
system with the span loss so large and the input optical power so
high, the output signals would be distorted heavily and become
unusable, were there no nonlinearity compensation
\cite{Wei03_lanl1}. By contrast, the transmission system becomes
virtually penalty-free with our scheme of mirror-symmetric
nonlinearity compensation \cite{Wei03_lanl1}. The test system in
translational symmetry is constructed with ten $100$-km spans on
one side of the OPC using the same transmission fiber as in the
mirror setup. Each span is ended by an erbium-doped fiber
amplifier (EDFA) with $20$ dB gain, noise figure $4$ dB, and a
dispersion compensating module (DCM) with negligible nonlinearity.
The DCM perfectly compensates the $D$ and $S$ of the fiber span.
On the other side of the OPC are ten spans of transmission fiber
with parameters $(\alpha',D',S',A'_{\rm
eff},\gamma',g')=(\alpha,-D,S,A_{\rm eff},\gamma,g)$. The loss and
dispersion of each span are also fully compensated by an EDFA and
a DCM. The EDFA noise figure is still $4$ dB. The input RZ pulses
peak at 10 mW when entering the transmission fiber. Firstly, the
OPC is absent and the transmission result of the 20-span link is
shown in the left-side graph of Fig.\ref{simu4transl}. When the
OPC is put back, the other graph in Fig.\ref{simu4transl} clearly
demonstrates the effect of nonlinearity compensation.

\begin{thebibliography}{}
\bibitem{Forghieri97} F. Forghieri, R. W. Tkach and A. R.
Chraplyvy, ``Fiber nonlinearities and their impact on transmission
systems,'' in {\em Optical Fiber Telecommunications III A}, I. P.
Kaminow and T. L. Koch, eds. Academic Press: San Diego, 1997.
\bibitem{Mitra01} P. P. Mitra and J. B. Stark, ``Nonlinear limits
to the information capacity of optical fiber communications,''
{\em Nature}, vol. 411, pp. 1027-1030, June 2001.
\bibitem{Watanabe96} S. Watanabe and M. Shirasaki, ``Exact
compensation for both chromatic dispersion and Kerr effect in a
transmission fiber using optical phase conjugation,'' {\em J.
Lightwave Techn.}, vol. 14, no. 3, pp. 243-248, 1996.
\bibitem{Brener00} I. Brener, B. Mikkelsen, K. Rottwitt, W. Burkett,
G. Raybon, J. B. Stark, K. Parameswaran, M. H. Chou, M. M. Fejer,
E. E. Chaban, R. Harel, D. L. Philen, and S. Kosinski,
``Cancellation of all Kerr nonlinearities in long fiber spans
using a LiNbO$_3$ phase conjugator and Raman amplification,'' {\em
OFC'00}, post-deadline paper, PD33, Baltimore, Maryland, 2000.
\bibitem{Wei03really} H. Wei and D. V. Plant, ``Does fiber
nonlinearity really limit the capacity of optical channels?'' {\em
OFC'03}.
\bibitem{Shen84} Y. R. Shen, {\em The Principles of Nonlinear
Optics}. New York: John Wiley \& Sons, 1984.
\bibitem{Agrawal95} G. P. Agrawal, {\em Nonlinear Fiber Optics},
2nd ed. San Diego: Academic Press, 1995.
\bibitem{Wei03_lanl1} H. Wei and D. V. Plant, ``On the Capacity of
Nonlinear Fiber Channels,'' arXiv:physics/0307020 at
http://arxiv.org/, July 2003.
\end{thebibliography}
\end{document}